\documentclass[preprint,aip,apl]{revtex4-1}

\usepackage{amsmath}    
\usepackage{graphicx}   
\usepackage{verbatim}   
\usepackage{color}      
\usepackage{epstopdf}  
\usepackage{upgreek} 
\usepackage[hidelinks]{hyperref}   
\usepackage{textcomp} 

\newcommand{\comsol}{COMSOL\texttrademark~}

\begin{document}

\title{Optimizing propagating spin wave spectroscopy}
\author{Juriaan Lucassen}
\email{j.lucassen@tue.nl}
\affiliation{Department of Applied Physics, Eindhoven University of Technology, 5600 MB Eindhoven, the Netherlands}
\author{Casper F. Schippers}
\affiliation{Department of Applied Physics, Eindhoven University of Technology, 5600 MB Eindhoven, the Netherlands}
\author{Luuk Rutten}
\affiliation{Department of Applied Physics, Eindhoven University of Technology, 5600 MB Eindhoven, the Netherlands}
\author{Rembert A. Duine}
\affiliation{Department of Applied Physics, Eindhoven University of Technology, 5600 MB Eindhoven, the Netherlands}
\affiliation{Institute for Theoretical Physics, Utrecht University, Leuvenlaan 4, 3584 CE Utrecht, the Netherlands}
\author{Henk J.M. Swagten}
\affiliation{Department of Applied Physics, Eindhoven University of Technology, 5600 MB Eindhoven, the Netherlands}
\author{Bert Koopmans}
\affiliation{Department of Applied Physics, Eindhoven University of Technology, 5600 MB Eindhoven, the Netherlands}
\author{Reinoud Lavrijsen}
\affiliation{Department of Applied Physics, Eindhoven University of Technology, 5600 MB Eindhoven, the Netherlands}

\date{\today}

\begin{abstract}
The frequency difference between two oppositely propagating spin waves can be used to probe several interesting magnetic properties, such as the Dzyaloshinkii-Moriya interaction (DMI). Propagating spin wave spectroscopy is a technique that is very sensitive to this frequency difference. Here we show several elements that are important to optimize devices for such a measurement. We demonstrate that for wide magnetic strips there is a need for de-embedding. Additionally, for these wide strips there is a large parasitic antenna-antenna coupling that obfuscates any spin wave transmission signal, which is remedied by moving to smaller strips. The conventional antenna design excites spin waves with two different wave vectors. As the magnetic layers become thinner, the resulting resonances move closer together and become very difficult to disentangle. In the last part we therefore propose and verify a new antenna design that excites spin waves with only one wave vector. We suggest to use this antenna design to measure the DMI in thin magnetic layers. 
\end{abstract}
\maketitle
Spin waves can be used to probe fundamental magnetic interactions in a ferromagnet. For example, the uniform spin wave mode is routinely used to determine the magnetic anisotropy in ferromagnetic resonance based techniques.~\cite{PhysRev.73.155,MAKSYMOV2015253} More recent advances have demonstrated that spin waves can further be used to probe spin polarized transport~\cite{doppler,PhysRevB.81.140407,doi:10.1063/1.4962835} and they are also frequently used to quantify the Dzyaloshinkii-Moriya interaction (DMI).~\cite{Lee2016,Nembach2015,Cho2015,PhysRevLett.114.047201} The use of spin waves to measure the DMI is especially interesting, because the field of skyrmionics revolves around this DMI.~\cite{Fert2017} Spin waves are one of the few ways of quantifying this interaction.~\cite{Ryu2013,PhysRevB.88.214401,Lee2016,Nembach2015,Cho2015,PhysRevLett.114.047201,doi:10.1021/acs.nanolett.6b01593} Measuring the DMI using spin waves utilizes the frequency difference for oppositely propagating spin waves as a direct result of the DMI.~\cite{PhysRevB.88.184404,0953-8984-25-15-156001} The most commonly used method to measure this frequency difference is Brillouin light scattering.~\cite{Nembach2015,Cho2015,PhysRevLett.114.047201}

Here we focus on the related, though less developed, technique of propagating spin wave spectroscopy (PSWS)~\cite{doppler} that can also measure the DMI induced frequency difference.~\cite{Lee2016,PhysRevB.93.235131} In PSWS, a micron sized coplanar waveguide is used to electrically generate spin waves with a specific wavevector in a magnetic strip via Oersted fields. These spin waves propagate towards a second antenna, where the spin waves are detected inductively. Although in principle PSWS is very sensitive to frequency differences, the fabrication of the devices is involved, and important details that are critical to correct operation remain underreported.

In this letter we demonstrate that the width of the magnetic strip critically determines the functionality of the device, with narrow strips being                                                                                                                                                                                                                                                                                                                                                                                                                                                                                                                                                                                                                                                                                                                                                                                                                                                                                                                                                                                                                                                                   optimal. First, we show that correcting for finite length of the microwave contacts (de-embedding) becomes important as the strip width increases. Second, for narrow strips, the spectra show additional resonances that belong to spin wave quantization modes along the strip width. Third, upon increasing the strip width we additionally find that the antenna-antenna coupling also increases, which detrimentally affects the spin wave transmission measurements. Last, we show a new antenna design which is truly monochromatic. This should aid the determination of DMI in magnetic films, as it allows the measurements to be performed for decreased strip thicknesses where the DMI is higher. Moreover, magnonic applications that require the presence of truly monochromatic spin waves can also benefit from this design.

\begin{figure}
\centering
\includegraphics{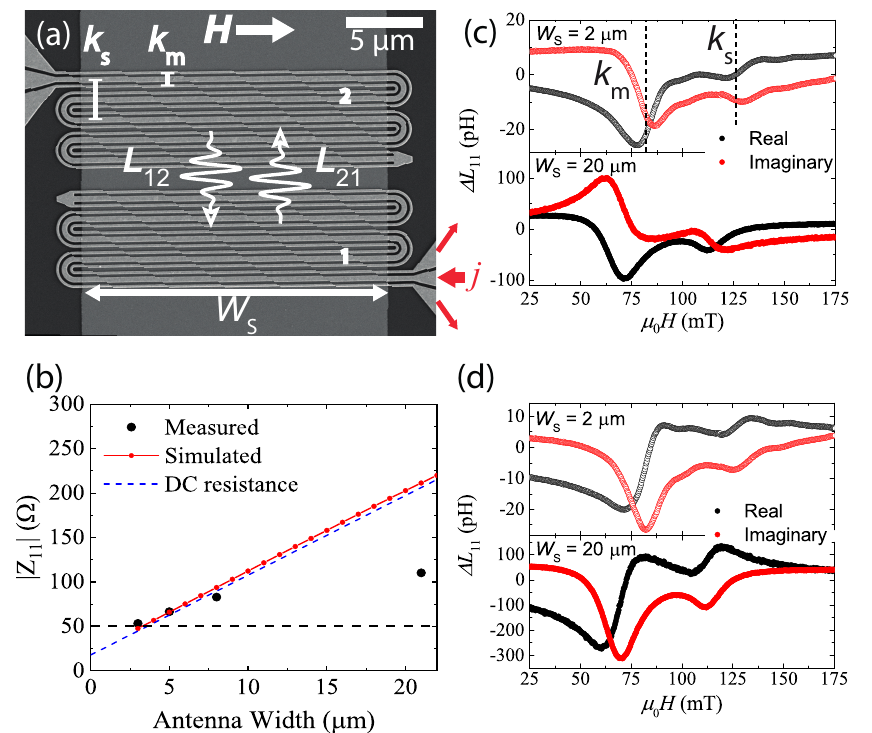}
\caption{\label{fig:figure1}(a) SEM micrograph of a fabricated device with $k_{\mathrm{m}}=7$~$\mathrm{\upmu}$m$^{-1}$ for a strip width $W_\mathrm{S}$ of $20$~$\mathrm{\upmu}$m. Also indicated is the direction of the magnetic field $H$, alternating current ($j$) flow  directions, main periodicities of the antenna, and antenna number. $L_{\mathrm{xy}}$ indicates the spin wave flow direction and corresponding mutual induction. (b) Measured absolute antenna impedances $Z_{\mathrm{11}}$ as a function of the antenna width at 14~GHz. Plotted together with the simulated antenna impedances as well as the calculated theoretical DC resistance of the antenna~(see supplementary material). The dashed horizontal line indicates $Z_\mathrm{0}=50~\Upomega$. (c) Raw $\Delta L_{\mathrm{11}}$ data at $14$~GHz with $k_{\mathrm{m}}=7$~$\mathrm{\upmu}$m$^{-1}$ for $W_\mathrm{S}=2$~$\mathrm{\upmu}$m~(top) and $20$~$\mathrm{\upmu}$m~(bottom). The two peaks, indicated by the dashed lines, correspond to the main periodicities of the antenna [see (a)]. (d) De-embedded version of data shown in (c).}
\end{figure}

We fabricated devices such as the one displayed in Fig.~\ref{fig:figure1}a. The operating principle of such a device is described in detail elsewhere.~\cite{PhysRevB.81.014425} In short, as we indicate in red in the figure, we drive a microwave current $j$ through one of the antennas. The spatial periodicity of the Oersted fields that couple to the spin waves is determined by the geometry of the antenna. Because there are two main periodicities, indicated by $k_{\mathrm{m}}$ and $k_{\mathrm{s}}$ in the figure, we also excite spin waves with these wave vectors. Spin waves then traverse the strip to the other antenna, where induction allows the spin waves to be detected. The magnetic strip underneath the antenna is fabricated using sputtering and an EBL lift-off process. The sputtered stack is //Ta(4)/Pt(4)/Co(15)/Ir(4)/Pt(4) (thicknesses in parentheses in nm) and was sputter deposited using Ar at $1 \times 10^{-2}$~mbar on a Si substrate with a native oxide in a system with a base pressure of $2 \times 10^{-9}$~mbar. On top of the magnetic strip, we deposited 40 nm of Al$_2$O$_3$ using ALD. Finally, the antennas were created using e-beam evaporation of Ti(10)/Au(100) in a second EBL lift-off process. We performed the spin wave resonance measurements using a VNA (Anritsu MS4644B) which we contacted to the antennas using microwave probes. The whole setup was calibrated using a microwave probe calibration substrate. Measurements were performed in field sweep mode with the magnetic field $H$ applied transverse to the strip, working in the Damon-Eshbach geometry at a power of $0$~dBm. Afterwards, the measured $S$ parameters were converted to inductions using well-known microwave relationships.~\cite{pozar2011microwave} Devices were fabricated for various strip widths $W_\mathrm{S}$ (2-20~$\upmu$m where the antenna width includes an additional $0.5~\upmu$m on each side) and antennas that were designed to excite different wave vectors ($k_\mathrm{m}=5~\mathrm{to}~9$~$\upmu$m$^{-1}$).

We start by looking at a typical measurement of the self-induction $\Delta L_{\mathrm{11}}$ as shown in Fig.~\ref{fig:figure1}c for two different $W_\mathrm{S}$. These measurements correspond to the amplitude of the spin waves that are excited by antenna $1$. We note two different peaks, indicated by the dashed lines, at two different fields which correspond to the two ($k_\mathrm{m}$ and $k_\mathrm{s}$) wave vectors of spin waves that are excited (later verified by fitting dispersion relation). Additionally, the curves resemble the (anti-)symmetric Lorentzian line shapes typical of ferromagnetic resonance for both strip widths. The phase for $W_\mathrm{S}=2$~$\upmu$m matches what one would expect for magnetic resonance: a symmetric imaginary induction and anti-symmetric real induction.~\cite{MAKSYMOV2015253} However, the phase of the $W_\mathrm{S}=20$~$\upmu$m device behaves rather differently, where the roles of the real and imaginary parts are now interchanged.

In Fig.~\ref{fig:figure1}d we plot the measurements of the self-induction corrected for the small change in the phase of the $S$ parameters as a result of the finite distance between the probes and the actual spin wave antenna. This process is called de-embedding.~\cite{pozar2011microwave} For $W_\mathrm{S}=2~\upmu$m, there is very little effect of de-embedding. However, for $W_\mathrm{S}=20~\upmu$m the phase of the spin wave resonances changes drastically and now matches the $W_\mathrm{S}=2~\upmu$m data. This is a rather surprising result because the induced phase difference $\theta$ as a result of the finite distance is only $\sim~40^{\circ}$ at $15$~GHz. Additionally, de-embedding only seems to be important for wider strips. To understand this behaviour, we derive the following relationship (with $\theta \ll 1$) for a 1-port circuit~\cite{pozar2011microwave}
\begin{equation}
\label{eq:deembed}
\Delta L_{\mathrm{11}} \rightarrow \Delta L_{\mathrm{11}}^{*} (1+\frac{i \theta Z_{\mathrm{11}}}{Z_{\mathrm{0}}}),
\end{equation}
where $\Delta L_{\mathrm{11}}$ is the proper de-embedded self-induction and $\Delta L_{\mathrm{11}}^{*}$ the measured self-induction. $Z_{\mathrm{11}}$ is the non-magnetic part of the impedance of the antenna and  $Z_{\mathrm{0}}$ is the characteristic impedance of the line (50~$\Upomega$). From this it is clear that de-embedding becomes more important as $Z_{\mathrm{11}}$ increases. In~Fig.~\ref{fig:figure1}b $|Z_{\mathrm{11}}|$ is plotted as a function of antenna width. It increases linearly with the antenna width, which explains why there is much larger effect of de-embedding for larger $W_\mathrm{S}$. This linear increase can be understood very simply in terms of the DC resistance of the antenna which dominates the impedance of the antennas (see supplementary material). 

Additionally, we see in Fig.~\ref{fig:figure1}c that the magnitude of the induction is only about 5 times larger for $W_\mathrm{S}=20~\upmu$m compared to $W_\mathrm{S}=2~\upmu$m. The induction should scale linearly with the magnetic volume, which is exactly what is found in Fig.~\ref{fig:figure1}d: a 10-fold increase in the induction going from the $2~\upmu$m strip to the  $20~\upmu$m strip. Once again, this can be understood from eq.~\eqref{eq:deembed}; there is not only a phase rotation present, but also a multiplicative term proportional to $Z_{\mathrm{11}}$. Although moving to smaller $W_\mathrm{S}$ will help decrease $Z_{\mathrm{11}}$ and thus remove the need for de-embedding, something similar can be achieved by decreasing the resistance of the antenna. For example, one can increase the thickness of the Au.~\footnote{This is no longer useful for thicknesses larger than the skin depth.}

\begin{figure}
\centering
\includegraphics{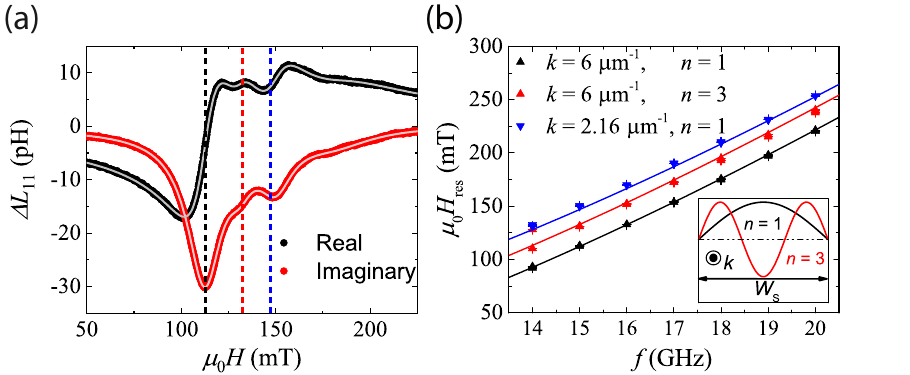}
\caption{\label{fig:figure2} Data with $k_{\mathrm{m}}=6$~$\mathrm{\upmu}$m$^{-1}$ at $W_\mathrm{S}=2~\upmu$m. (a) $\Delta L_{\mathrm{11}}$ at $15$~GHz plotted together with a fit of the 3 spin wave resonances observed. The vertical dashed lines indicate the resonance fields $H_{\mathrm{res}}$ obtained from the fit. (b) Fitted resonance fields $H_{\mathrm{res}}$ as a function of frequency $f$ [see (a)]. In the inset we show schematically the lateral (along the strip width) spin wave quantization modes that are used to fit the resonance fields.}
\end{figure}

Next, we demonstrate that upon decreasing $W_\mathrm{S}$, a spin wave quantization resonance appears in the spectra. To see this more clearly, we plot $L_{\mathrm{11}}$ data for a $W_\mathrm{S}=2~\upmu$m strip in Fig.~\ref{fig:figure2}a. Once again note that there are two main peaks present in this figure; the $k_{\mathrm{m}}$ peak at $\sim 110$~mT and the $k_{\mathrm{s}}$ peak at $\sim 150$~mT, but there is clearly another resonance visible at $\sim 140$~mT. This resonance vanishes as $W_\mathrm{S}$ is increased to $20~\upmu$m. From this we conclude that any additional periodicities of the antenna geometry that can couple to this spin wave can be excluded, because then it should be present for both $W_\mathrm{S}=2$~and~$20~\upmu$m devices.~\cite{PhysRevB.81.014425} Instead, we believe it to be a higher order laterally quantized spin wave mode (inset~Fig.~\ref{fig:figure2}b), which to our knowledge has not yet been reported in PSWS measurements.

A more detailed quantitative analysis can be performed by fitting the dispersion relation to the resonance fields $H_{\mathrm{res}}$. We obtain these $H_{\mathrm{res}}$ by fitting the spectrum of Fig.~\ref{fig:figure2}a to a combination of symmetric and anti-symmetric Lorentzian lineshapes.~\footnote{We use 4 symmetric- and anti-symmetric Lorenzians to fit the curves. We need a fourth curve to properly fit the background for some measurements. We believe there to be a fourth spin wave resonance - the quantization of the $k_{s}$ peak - at higher fields, but we are not able to reliable fit this peak.} The resulting $H_{\mathrm{res}}$ are indicated by the vertical dashed lines. For all three resonances, $H_{\mathrm{res}}$ is plotted in Fig.~\ref{fig:figure2}b as a function of the frequency $f$. 

These curves are fitted simultaneously using dispersion relations derived elsewhere~\cite{Kalinikos1981,0022-3719-19-35-014} which are also plotted in Fig.~\ref{fig:figure2}b. Here, we use $g=2.17$, $M_{\mathrm{s}}=1.44$~MA~m$^{-1}$ and $k_{\mathrm{s,m}}=2.16,6$~$\upmu$m$^{-1}$ (fixed by the antenna geometry). We assume that the quantized spin wave mode is an $n=3$ mode (mode profiles are indicated in the inset of Fig.~\ref{fig:figure2}b) because the excitation efficiency for the $n=2$ mode is negligible.~\cite{PhysRev.110.1295} The quantization is taken into account by adding a wavevector $k=\frac{n\pi}{W_\mathrm{S}}$ perpendicular to the propagation direction in the dispersion relation.~\cite{PhysRevB.66.132402} We use the following fit parameters: an effective strip width $w_{\mathrm{eff}}$, layer thickness $t$ and $M_\mathrm{eff}=M_\mathrm{s}-H_\mathrm{K}$, with $H_\mathrm{K}$ the magnetic anisotropy field. The resulting fit gives $M_{\mathrm{eff}}=1.1\pm 0.1$~MA~m$^{-1}$, $w_{\mathrm{eff}}=1.1\pm 0.8$~$\upmu$m, and $t=11\pm 4$~nm. $M_{\mathrm{eff}}$ is reasonable for this system.~\cite{Lee2016} Because we do not take into account the non-uniform internal dipolar fields,~\cite{PhysRevB.66.132402,PhysRevB.68.024422} the underestimation of $W_{\mathrm{S}}$ and $t$ is not surprising. In the supplementary material we present fits for devices with different $k_{\mathrm{m}}$ values. 

\begin{figure}
\centering
\includegraphics{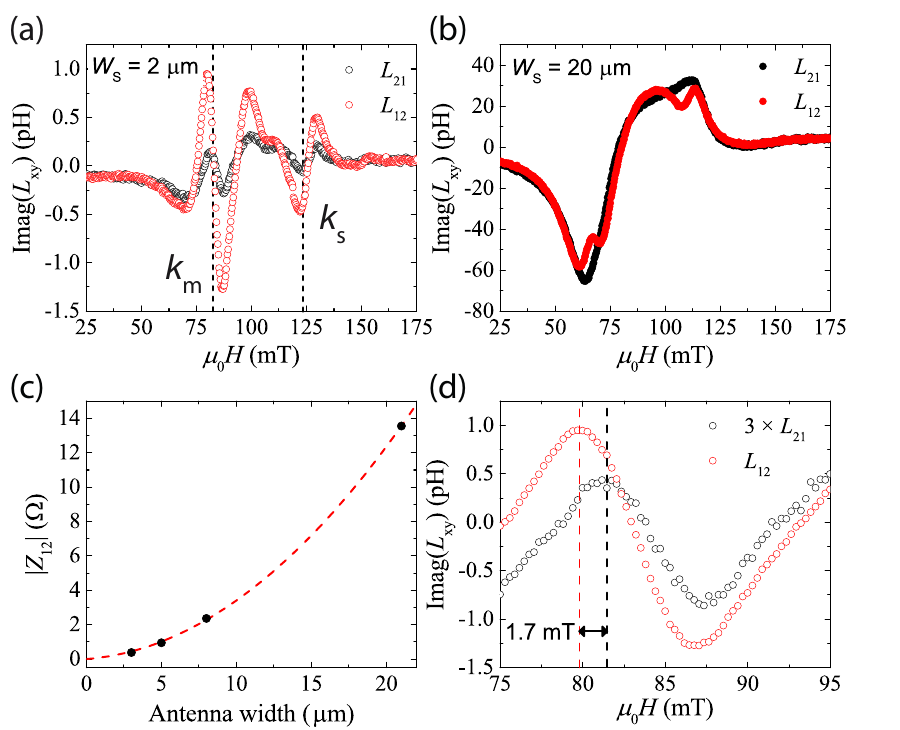}
\caption{\label{fig:figure3} $\mathrm{Imag(}\Delta L_{\mathrm{xy}}\mathrm{)}$ data at $14$~GHz with $k_{\mathrm{m}}=7$~$\mathrm{\upmu}$m$^{-1}$ for $W_\mathrm{S}=2~\upmu$m~(a) and $20$~$\mathrm{\upmu}$m~(b). The two dashed lines in (a) indicate the resonance fields of spin waves corresponding to the main periodicties of the antenna (see Fig.~\ref{fig:figure1}a). (c) Absolute value of the antenna-antenna coupling impedance $Z_\mathrm{12}$ as a function of the antenna width. The line is a guide to the eye, based on a quadratic fit. (d) Zoomed-in version of (a) with the peak-shift of $\sim 1.7$~mT indicated. Here, $L_{\mathrm{21}}$ was artificially blown up to make the peak shift easier to see.}
\end{figure}

We now turn our attention to the spin wave transmission measurements. A typical measurement for $W_\mathrm{S}=2~\upmu$m is plotted in Fig.~\ref{fig:figure3}a, where we plot the mutual induction $\Delta L_{\mathrm{12}}$ ($\Delta L_{\mathrm{21}}$) which corresponds to spin waves traveling from antenna 2 (1) to 1 (2) (see Fig.~\ref{fig:figure1}a). Once again, we can distinguish two peaks corresponding to the two different type of spin waves that are excited.  Note two very distinct features that are indicative of a proper electrical spin wave transmission signal: first, a distinct amplitude asymmetry between oppositely traveling spin waves ($L_{\mathrm{12}}$ vs $L_{\mathrm{21}}$), which is the result of the chirality of the driving fields that matches the corresponding spin wave ($L_{\mathrm{12}}$) or opposes it ($L_{\mathrm{21}}$).~\cite{PhysRevB.77.214411} Second, sharp oscillations of the spin wave transmission signal which are the result of a variation in the spin wave phase as we sweep through the resonance.~\cite{PhysRevB.81.014425} 

However, a similar measurement for $W_\mathrm{S}=20~\upmu$m yields Fig.~\ref{fig:figure3}b. Both the amplitude asymmetry and the sharp oscillations now no longer seem present. This is rather surprising, as both features have an origin that does not depend on $W_\mathrm{S}$. Rather, we believe it is related to a direct parasitic coupling between the two antennas. This means that if a spin wave is excited by antenna 1 there is a signal induced in antenna 2 independent of an actual physical spin wave being transmitted.~\footnote{A simple way to check this would entail removing the magnetic strip between the antennas only. However, as detailed in the supplementary material, this parasitic coupling is also mediated by the magnetic strip. Therefore, it still remains to be explicitly verified that the parasitic coupling is independent of spin waves being transmitted.} For example, for $L_{\mathrm{12}}$ there is still a small oscillatory signal superimposed on the large resonant background. This background is the result of the parasitic coupling and the small superimposed signal is the transmitted spin wave. The spin wave transmission signal for $L_{\mathrm{21}}$ is smaller, as observed in Fig.~\ref{fig:figure3}a, such that the smaller oscillatory signal on top of this induction is no longer visible in Fig.~\ref{fig:figure3}b.

The magnitude of the parasitic coupling $|Z_{\mathrm{12}}|$ is plotted in Fig.~\ref{fig:figure3}c, where we find that the coupling seems to scale quadratically with the antenna width. This explains why devices with smaller $W_\mathrm{S}$ do show a proper spin wave transmission signal. Yet, even for small $W_\mathrm{S}$ this parasitic coupling can become problematic at higher frequencies where the increasing spin wave attenuation decreases the spin wave transmission signal.~\footnote{Moving the antennas closer together should remedy (at least part of) this issue} At present, we cannot explain the size and behaviour of this coupling, but more details can be found in the supplementary material.

\begin{figure}
\centering
\includegraphics{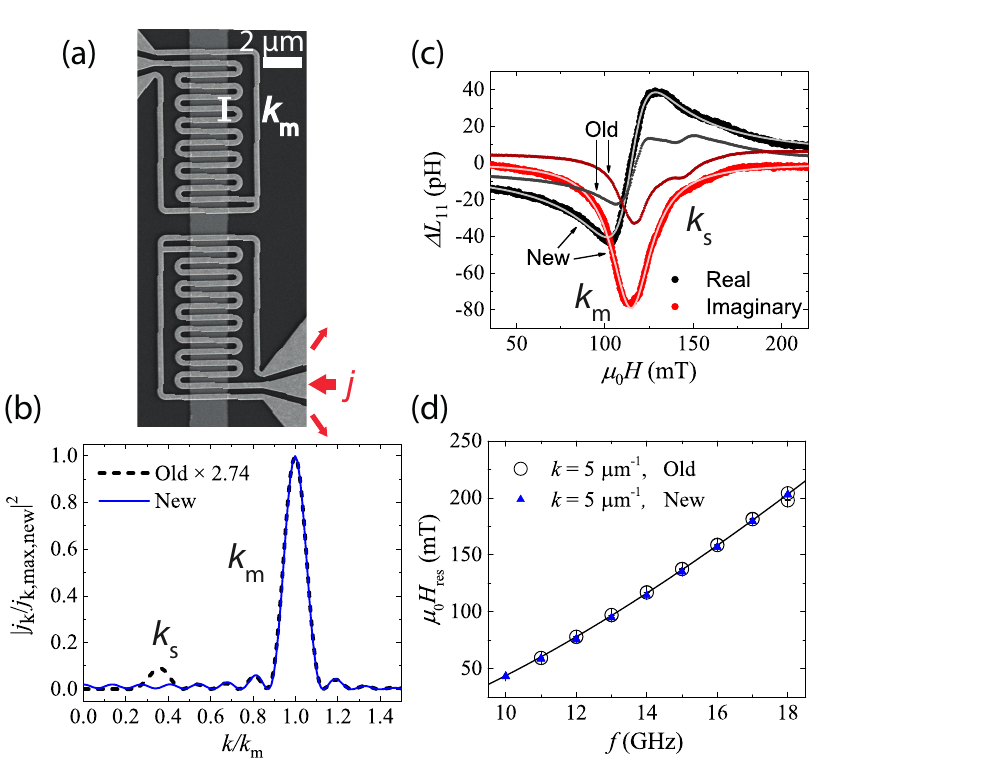}
\caption{\label{fig:figure4}(a) SEM micrograph of a fabricated device with $k_{\mathrm{m}}=5$~$\mathrm{\upmu}$m$^{-1}$ and $W_\mathrm{S}=2~\upmu$m. Also indicated are the alternating current ($j$) flow directions and main periodicity of the antenna. (b) Square of the Fourier transform of the current density $j_{\mathrm{k}}$ of the old and new antenna design. Notice the scaling of the old antenna design. (c) $\Delta L_{\mathrm{11}}$ at $14$~GHz with $k_{\mathrm{m}}=5$~$\mathrm{\upmu}$m$^{-1}$ for $W_\mathrm{S}=2~\upmu$m for both an old and new antenna design. For the new design, the solid line is the result of a fit. (d) Resonance fields $H_{\mathrm{res}}$ as a function of frequency $f$ at $k_{\mathrm{m}}=5$~$\mathrm{\upmu}$m$^{-1}$ and $n=1$ for both the old and new design. The fit belongs to the complete data set of the old design.~\cite{fit}}
\end{figure}

For the $W_\mathrm{S}=2~\upmu$m device a peak shift can be extracted that could be a measure for the DMI. This shift is shown in Fig.~\ref{fig:figure4}d, where $L_{\mathrm{12}}$ is shifted about $+1.7$~mT with respect to $L_{\mathrm{21}}$. The shift is opposite to the direction expected from DMI [assuming $D_{\mathrm{s}}=1.8$~pJ~m$^{-1}$~(Ref.~\onlinecite{doi:10.1021/acs.nanolett.6b01593})], which is about $-1.7$~mT. This shift can have other contributions beyond the DMI, such as the anisotropy difference induced shift.~\cite{PhysRevB.96.174420} Upon moving to thinner layers, this contribution should decrease in size, and the contribution of the DMI to the field shift will increase. Therefore, in future work, we would like to investigate thinner layers to determine the origin of this shift. 

In the final part of this letter, we a present a new antenna design. This has the major advantage of exciting only one type spin wave ($k_\mathrm{m}$) which is necessary if thinner layers have to be investigated. Upon decreasing the layer thickness the two traditional spin wave resonances (corresponding to $k_{\mathrm{m}}$ and $k_{\mathrm{s}}$) start overlapping  because the interfacial anisotropy term becomes more important and because of the decreasing influence of the magnetostatic interactions. Although the ratio between the $k_\mathrm{m}$ and $k_\mathrm{s}$ resonance is quite large for $L_\mathrm{11}$, they are of approximately equal size in the transmission measurement as seen in Fig.~\ref{fig:figure3}a. If the two peaks move closer together, disentangling the two resonances becomes increasingly difficult in a transmission measurement.

The new antenna design is shown in Fig.~\ref{fig:figure4}a. Rather than relying on a conventional CPW signal and ground line structure, in this new design only the signal line is meandered. There is no need to adhere to conventional coplanar waveguide structures for these spin wave antennas as the antennas are much smaller than the electrical wavelength. In the figure we indicate the only periodicity $k_{\mathrm{m}}$ present such that spin waves with only one wave vector are excited. This should negate the problem of overlapping spin wave resonances in the transmission induction spectra. To further illustrate how this works, note that in Ref.~\onlinecite{PhysRevB.81.014425} it is demonstrated that the spin wave excitation signals are proportional to the square of the spatial Fourier transform of the current density used to excite the spin waves. In Fig.~\ref{fig:figure4}b the Fourier transform of the current density for both the old (Fig.~\ref{fig:figure1}a) and new (Fig.~\ref{fig:figure4}a) antenna design are plotted. For the old design, there are two peaks ($k_{\mathrm{s}}$ and $k_{\mathrm{m}}$) that correspond to the two spin wave resonances that are measured. For the new design, however, the secondary peak at $k_{\mathrm{s}}$ disappears, meaning that with this new design spin waves with only one wave vector $k_{\mathrm{m}}$ are excited. Moreover, the $k_\mathrm{m}$ peak of the new design is $\sim~3$ times larger than the old design as a result of the higher current density that flows through the new design, suggesting the induction signals should also be larger.

We verify these predictions by measuring the self-induction $\Delta L_{\mathrm{11}}$ for the new antenna design; this is plotted in Fig.~\ref{fig:figure4}c together with a similar measurement on a device with the old antenna design. As can be seen, the secondary peak at $k_\mathrm{s}$ has vanished for the new antenna design, agreeing with our initial expectations based on the periodicity of the antenna. The intensity of the signal is also a factor $\sim~2$ larger which agrees mostly with the initial predictions based on the current density.

A more thorough analysis is obtained by fitting the spectra to obtain the resonance fields $H_{\mathrm{res}}$. Such a fit is also displayed in Fig.~\ref{fig:figure4}c with solid lines.~\footnote{ We use only 1 symmetric and anti-symmetric Lorentzian to fit the spectrum. The higher order mode for the new design is not fitted. It is visible but the fits consistently placed the peak at a different location.} Combining this with a dispersion relation analysis similar to the one performed in Fig.~\ref{fig:figure2}b yields Fig.~\ref{fig:figure4}d, where we plot only the main resonance field of the spectra. The resonance fields for the new design, shown in blue, lay perfectly on top of the data of the old design.~\cite{fit}

To summarize, we have demonstrated the benefit of using narrower strips for propagating spin wave spectroscopy (PSWS). We ended the letter with a demonstration of a new antenna design that allowed us to excite spin waves with only one wave vector suitable for the investigation of DMI in thinner films.

See supplementary materials for (1) details on the \comsol simulations, (2) additional $Z_\mathrm{11}$ information, (3) a full dispersion relation fit and (4) additional details on the parasitic coupling $Z_\mathrm{21}$.

This work is part of the research programme of the Foundation for Fundamental Research on Matter (FOM), which is part of the Netherlands Organisation for Scientific Research (NWO).
\bibliography{references}

\begin{thebibliography}{31}
\bibitem [{fit()}]{fit}%
  \BibitemOpen
  \href@noop {} {}\bibinfo {note} {We fitted all resonance fields, similar to
  the analysis in Fig.~\ref{fig:figure2}b, using $g=2.17$,
  $M_{\mathrm{s}}=1.44$~MA~m$^{-1}$ and
  $k_{\mathrm{s,m}}=2.16,6$~$\upmu$m$^{-1}$ the fit finds
  $M_{\mathrm{eff}}=1.02\pm 0.01$~MA~m$^{-1}$, $w_{\mathrm{eff}}=1.2\pm
  0.1$~$\upmu$m, and $t=10\pm 0.6$~nm. These values agree well with the data
  from Fig.~\ref{fig:figure2}b. The fit uncertainties for the quantization and
  secondary peak are quite high - to avoid getting non-nonsensical
  uncertainties we therefore normalize the fit uncertainties by
  $\chi^2$.}\BibitemShut {Stop}%
\end{thebibliography}
\end{document}